\documentclass[a4paper,11pt]{article}

\usepackage{pos}
\usepackage{caption}
\usepackage{subcaption}
\usepackage{cleveref}
\usepackage{multirow,booktabs,xfrac}
\usepackage{tikz}           

\usepackage{personal-macros}
\newcommand{\scG}{\phi}
\newcommand{\scF}{\varphi} 
\newcommand{\scA}{\Sigma} 
\newcommand{\fluc}{\eta} 

\title{Exploring $\SU[\text{3}]$-Higgs theories} 

\author*{Elizabeth Dobson}
\author*{Axel Maas}
\author*{Bernd Riederer}

\emailAdd{elizabeth.dobson@uni-graz.at}
\emailAdd{axel.maas@uni-graz.at}
\emailAdd{bernd.riederer@uni-graz.at}

\notes{\note[]{Combined Proceedings for the talks ``Exploring \SU[3] + Higgs theories: the fundamental case'' and ``Exploring \SU[3] + Higgs theories - The adjoint case''}}

\affiliation{Institute of Physics, NAWI Graz, University of Graz,\\
  Universitätsplatz 5, Graz, Austria}

\abstract{
	The requirement of manifest gauge invariance leads to a conflict between perturbative and non-perturbative predictions for the low-energy spectra of grand-unified theories. These conflicts already emerge in simplified prototype models of \SU[3] gauge theories with Higgs fields in different representations. We expand earlier lattice investigations on this subject and provide further support for the predicted deviations. These can be understood in terms of the Fr\"ohlich-Morchio-Strocchi mechanism.
}

\FullConference{%
 The 38th International Symposium on Lattice Field Theory, LATTICE2021
  26th-30th July, 2021
  Zoom/Gather@Massachusetts Institute of Technology
}

\begin{document}
\maketitle

\section{Introduction}

Grand-unified theories (GUTs) \cite{Langacker:1980js} are an attractive explanation for the, in the Standard Model unrelated, electric charges of the fermions. Furthermore, they appear to elegantly continue the unification of forces, which started with electromagnetism.
However, the usual approach to them relies, as in the Standard Model, on a spontaneous breaking of the gauge symmetry. Such a breaking is forbidden by Elitzur's theorem \cite{elitzur1975}. In the Standard Model, the Fr\"ohlich-Morchio-Strocchi (FMS) mechanism \cite{frohlich1981} alleviates this discrepancy, and explains why, even when ignoring Elitzur's theorem, quantitatively little changes.

Unfortunately, it turns out that for generic GUTs this does not happen \cite{maas2019a,sondenheimer2020}: general aspects like the spectrum differ qualitatively from the one expected from a literal breaking of the gauge symmetry. This has been supported by exploratory lattice simulations \cite{maas2018,afferrante2020a}; for a review see \cite{maas2019}. This does not imply that the idea of GUTs does not work in principle. But it substantially alters the class of theories suitable to serve as GUTs, and especially excludes most popular candidates \cite{sondenheimer2020,maas2019a}. To be able to build more suitable candidates requires thus a firm understanding of suitable analytic tools. In fact, the FMS mechanism has again proven so far to be capable of such a description \cite{maas2019}, and it essentially takes the form of an augmentation of perturbation theory, thus applicable with comparable ease \cite{maas2020a}.

Still, there remain a couple of issues for which the FMS mechanism needs to be better understood, which have no analogue in the Standard Model. These relate to states, which have global quantum numbers under weakly-coupled interactions, which do not appear in the elementary spectrum \cite{maas2018,maas2019a}, or in cases with multiple conventional breaking patterns \cite{maas2019a}.

To ensure a reliable understanding requires a confirmation with first-principle, non-perturbative methods, such as lattice simulations. After all, the FMS mechanism is perturbative in nature \cite{maas2019}, even if it transcends ordinary perturbation theory.

We present here preliminary results for this endeavour. To this end, we study the simplest theories which exhibit the features in questions, both to avoid interference of more complex structure and to keep the necessary amount of computing manageable. These are \SU[3] Yang-Mills theories coupled to a single scalar field in either the fundamental or the adjoint representation. Our results corroborate previous exploratory findings, and determine the next steps.

\section{Gauge-scalar theories}

\subsection{Continuum formulation}

A general $\SU[N]$ gauge-scalar theory can be described by the Lagrangian
\begin{equation}\label{eqn:general_lagrangian}
        \cL = -\frac{1}{4}W_{\mu\nu}W^{\mu\nu}
        +\qty(D_\mu\scG^R)^\dag \qty(D^\mu\scG^R)
        - V\qty({\scG^{R}}^{\dag}\scG^R)\,,
        \quad\text{with}\;\;
        W^{\mu\nu}=\pd^\mu W^\nu-\pd^\nu W^\mu + ig \qty[W^\mu,W^\nu]\,,
\end{equation}
where $W^{\mu\nu}$ is the gauge-field strength tensor and $V\qty(\scG^\dag\scG)$ is a gauge invariant potential that allows for a Brout--Englert--Higgs (BEH) effect.
The label $R$ here refers to the representation of the scalar field -- i.e., fundamental or adjoint with respect to $\SU[N]$ -- and $D_\mu$ is a covariant derivative which acts on the scalar field as 
\begin{equation}
    D_\mu\scG^R = \qty(\pd_\mu-ig W^a_\mu T_a^R)\scG^R
    \quad\text{with generators }\qty(T_a)_{ij}^R=\;\;
    \begin{cases}
        \qty(t_a)_{ij} & R=\text{`fundamental'}
        \\
        -i f^{aij} & R=\text{`adjoint'}
    \end{cases}
\end{equation}
In the fundamental case, we use a vector representation of the scalar with $\scG^\text{Fun.}\equiv\scF$ and denote its (complex) components as $\scF_a$. 
Under a gauge transformation,
\begin{align}
\text{with }G(x)\in \SU[N]\,,\quad
  \scF(x)&\to \scF'(x)=G(x)\scF(x)\\
  \qty(D_\mu\scF)(x) = \qty(\pd_\mu\scF)(x) - i g W_\mu(x)\scF(x)
  &\to \qty(D_\mu\scF)'(x) = G(x)\qty(D_\mu\scF)(x)\,.
\end{align}
In the adjoint case we use a matrix representation of the algebra-valued scalar field instead,  $\scA(x)=\scA^a(x)T_a$ to make the difference clear.
Under a change of gauge, the field transforms as 
\begin{align}\label{eq:adj_field_descr}
	\scA(x)&\to \scA'(x)=G(x)\scA(x)G^\dag(x)\\
	\qty(D_\mu\scA)(x) = \qty(\pd_\mu\scA)(x) + i g\qty[W_\mu(x),\scA(x)]
	&\to \qty(D_\mu\scA)'(x) = G(x)\qty(D_\mu\scA)(x)G^\dag(x)\,.
\end{align}
We note that in the fundamental case there is an additional global \U[1] symmetry acting only on the scalar field, while in the adjoint case this is merely a \ZZ[2] symmetry. We will assume that the potential leaves these symmetries intact, and that it allows for a BEH effect at tree-level.\footnote{Note that in the adjoint case the most general potential would contain also a $\tr\qty[\scA^3]$ term which we explicitly exclude by this requirement.}
In the following we will concentrate on \SU[3] as the gauge group. Note that the global symmetry is the same for all \SU[N] gauge groups with $N\ge3$.

\subsection{Gauge symmetry and physics}

The usual approach to such theories \cite{Bohm:2001yx,maas2019a} is to implement the BEH-effect after gauge-fixing and split the Higgs field as $\scG(x)=v\scG_0+\fluc(x)$, where $v$ is the vacuum expectation value, $\scG_0$ is a unit vector fixed by the gauge choice, and $\fluc$ is the fluctuation field.
In the present case this corresponds in the fundamental case to a breaking of $\SU[3]\to\SU[2]$ while in the adjoint case two patterns, $\SU[2]\times\U[1]$ and $\U[1]\times\U[1]$, are possible. Depending on the realised breaking pattern this gives mass to 5/4/6 of the gauge bosons respectively, and leaves one massive and 0/3/1 massless scalar fields. The fields are multiplets or singlets with respect to the corresponding unbroken groups. Note that in the usual gauge choices an unbroken diagonal subgroup of the gauge symmetry and the global symmetry remains, under which some of the states are charged.

As noted, Elitzur's theorem forbids this, so the gauge symmetry needs to stay intact. Thus, only gauge-invariant states can be physical, which are necessarily composite. Such states can only be classified according to (unbroken) global groups. Since these are abelian (i.e. \U[1] and \ZZ[2]), their representations are one-dimensional, and thus no degeneracy patterns appear, except accidental ones.

As the same reasoning also applies to the Standard Model, this appears to be in contradiction with the wealth of experimental data. It is here, where the FMS mechanism comes into play. This approach takes the stance of Elitzur's theorem as a starting point, then uses only explicitly gauge-invariant quantities.

Consider, e.\ g., the uncharged scalar channel. A suitable gauge-invariant operator is necessarily a composite operator, e.\ g., $\scF^\dagger\scF$ or $\tr\qty[\scA^3]$. 
Since the BEH effect can be considered to be due to gauge-fixing alone, it is viable to work in a Landau--'t Hooft gauge. The operators can then be written as, e.\ g.,
\begin{equation}
    \scG^\dagger\scG=v^2+v\qty(\scG_0^\dagger\fluc+\fluc^\dagger \scG_0)+\fluc^\dagger\fluc
    \equiv v^2 + v h + \fluc^\dag\fluc\,,
\end{equation}
where $h=2\Re\qty(\scG_0^\dag\fluc)$ is the `elementary Higgs' field.
Ignoring the constant term, this implies that to leading order in $\fluc/v$ the composite operator behaves like the single-particle operator. When constructing gauge-invariant matrix elements, the combination of this additional expansion with the usual perturbative one constitutes the FMS mechanism. It is therefore an analytic and systematic approach within the validity of the expansions.

In particular it follows  for the propagator that
\begin{equation}
    \ev{\qty(\scG^\dag\scG)^{\dagger}(x)\qty(\scG^\dag\scG)(y)}_c \approx v^2\ev{h(x)^{\dagger}h(y)}_c
\end{equation}
to leading order, and thus the poles coincide. In the Standard Model, there is a one-to-one mapping between the poles of the composite particles and an elementary one in every channel, and the same is true for all matrix elements to leading order in $\fluc/v$. This explains why, in the Standard Model, the usual picture of the BEH effect coincides with the formally correct approach of the FMS expansion. This is also confirmed in lattice simulations: see \cite{afferrante2020a,maas2019} for an overview.

The reason for this is that the remaining global symmetry in the Standard Model and the weak gauge group (i.e. \SU[2]) coincide. Thus, for every gauge-dependent multiplet there exists a gauge-invariant multiplet, on which the masses and degeneracy pattern can be mapped. In general GUTs, including our simplified models, this is not the case. As a consequence, there is generically no one-to-one correspondence, and thus the spectrum differs qualitatively \cite{sondenheimer2020,maas2019a}. Moreover, if multiple breaking patterns are possible just by a choice of gauge, it is not even clear what should be mapped to the physical spectrum at all \cite{maas2019a}. We explore both possibilities in the following using lattice simulations to avoid performing an expansion, and will afterwards analyse to which extent the results can be interpreted, or even determined, using the FMS mechanism.

\subsection{Lattice formulation}

To this end, we employ lattice simulations. We perform simulations of the theory on a four-dimensional euclidean lattice of volume $V=L^4$. The discretization of the theory stated in \cref{eqn:general_lagrangian,eqn:adj_potential} is given by the following action
\begin{align}\label{eqn:adj_lattice_lagrangian}
    S= \sum_{x} & \beta\qty[1-\frac{1}{3}\sum_{\mu<\nu}\Re\qty[\tr\qty[U_{\mu\nu}\qty(x)]]]  + S_\scG^i\\
    S_\scG^F&=\gamma[\scF^\dagger\scF - 1]^2 +  \scF^\dagger \scF - 2\kappa\sum_{\mu=1}^{4} \scF^\dagger(x)U_{\mu}(x)\scF(x+\hat{\mu})\\
    S_\scG^A&=\gamma\qty[2\tr\qty[\scA\qty(x)^2] - 1]^2 +  2\tr\qty[\scA\qty(x)^2] - 2\kappa\sum_{\mu=1}^{4} \tr\qty[\scA\qty(x)U_{\mu}\qty(x)\scA\qty(x+\hat{\mu})U_{\mu}\qty(x)^{\dagger}]
\end{align}
with $\beta$ the gauge coupling, $\kappa$ the hoping parameter, $\gamma$ the quartic coupling and the standard conventions for the plaquette $U_{\mu\nu}\qty(x)$. The lattice parameters are related to the continuum parameters by $\beta = 6/g^2$, $a^2\mu^2=(2-4\gamma)/\kappa - 8$ and $\lambda=8\gamma/\kappa.$

The simulations used a heatbath algorithm for the link updates \cite{cabibbo1982}. In the fundamental case an exact heatbath algorithm can be achieved by a suitable modification of the staples to incorporate the interaction term. For the adjoint case an additional Metropolis step has been used to account for the interaction. The scalar updates were performed as a generalized pseudo-heatbath method like the one proposed in \cite{knechtli1999}. Here we solve the resulting cubic equation in this update \cite[eq. (C.43)]{knechtli1999} exactly, to obtain a higher acceptance rate in areas where $\kappa$ becomes large. A configuration is obtained after one full sweep, which consists of five pure link sweeps followed by one scalar sweep. For the adjoint case additional overrelaxation sweeps for both the links and the scalar fields have been performed. To ensure decorrelation, sufficient configurations have been dropped for thermalization andin between measurements have been dropped, yielding an autocorrelation time of 1 for local quantities like the plaquette. The results were obtained from a combination of many individual runs.

\section{The fundamental case}

\subsection{Predicted spectra}

\begin{figure}[h]
	\centering
    \begin{tikzpicture}[scale=\tikzscale, every node/.style={scale=\tikzscale}]
    \draw[->,line width=2pt] (0,1) -- (0,7);
    \draw[line width=2pt] (-0.08,1.2) -- (0.08,1.2);
    \node (A) at (-0.5,6.5) {\rotatebox{90}{mass}};
    \node (B) at (-0.4,1.2) {$0$};
    \draw[line width=2pt] (-0.08,3) -- (0.08,3);
    \node (BB) at (-0.3,3.2) {\scriptsize{$m_a$}};
    \draw[line width=2pt] (-0.08,4) -- (0.08,4);
    \node (BBB) at (-0.3,4) {\scriptsize{$m_A$}};
    \draw[line width=2pt] (-0.08,4.5) -- (0.08,4.5);
    \node (BBB) at (-0.3,4.5) {\scriptsize{$m_h^1$}};
    \draw[line width=2pt] (-0.08,5.2) -- (0.08,5.2);
    \node (BB) at (-0.4,5.2) {\scriptsize{$2m_a$}};
    
    \node (C) at (2.5-0.2,7) {Perturbation theory};
    \node (D) at (2.5-0.2-1,6.5) {scalar};
    \node (E) at (2.5-0.2+1,6.5) {vector};
    
    \fill[vector0] (1.55+0.3-0.2+1.2,1.2) circle (0.2);
    \fill[vector0] (2.05+0.3-0.2+1.2,1.2) circle (0.2);
    \fill[vector0] (2.55+0.3-0.2+1.2,1.2) circle (0.2);
    
    \fill[vector1] (1.3+0.3-0.2+1.2,3.2) circle (0.2);
    \fill[vector1] (1.8+0.3-0.2+1.2,3.2) circle (0.2);
    \fill[vector1] (2.3+0.3-0.2+1.2,3.2) circle (0.2);
    \fill[vector1] (2.8+0.3-0.2+1.2,3.2) circle (0.2);
    
    \fill[vector2] (2.05+0.3-0.2+1.2,4) circle (0.2);
    
    \fill[scalar] (2.5-0.2-1,4.5) circle (0.2);
    
    \node (F) at (7.5,7) {Gauge-invariant};
    \node (G) at (6.5-0.4,6.5) {U$(1)$-singlet};
    \node (H) at (6.5-1.1,6) {scalar};
    \fill[scalar] (6.5-1.1,4.5) circle (0.2);
    \node (I) at (6.5+0.3,6) {vector};
    \fill[vector2] (6.5+0.3,4) circle (0.2);
    \node (J) at (9.5-0.4,6.5) {U$(1)$-non-singlet};
    \node (H) at (9.5-0.35,6) {scalar \& vector};
    \fill[scalar] (9.5-0.6,5.2) circle (0.2);
    \fill[vector1] (9.5-0.6,5.2) -- +(90:0.2) arc (90:-90:0.2) -- cycle;
    \fill[scalar] (9.5-0.1,5.2) circle (0.2);
    \fill[vector1] (9.5-0.1,5.2)  -- +(90:0.2) arc (90:-90:0.2) -- cycle;
    
    \node (foot) at (8.5,1) {\tiny{$^1$May have any energy }};
\end{tikzpicture}
	\caption{The expected spectrum at tree level obtained through standard perturbation theory (left) differs qualitatively from that of the explicitly gauge invariant FMS approach (right).}
	\label{fig:fun_spectra}
\end{figure}
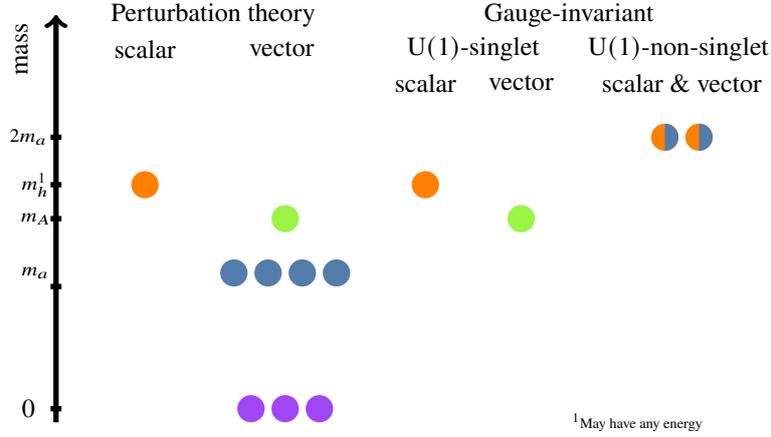

\Cref{fig:fun_spectra} shows the perturbative spectra. The mass of the remaining Higgs particle can be selected arbitrarily, and there are also three massless gauge bosons in the adjoint representation of the unbroken \SU[2] subgroup, two pairs of degenerate massive gauge bosons in the fundamental and anti-fundamental (which coincide) representation of \SU[2] with mass $m_a$, and a singlet of \SU[2] with mass $m_A=\sqrt{4/3}m_a$.

The physical spectrum, on the other hand, cannot be classified in terms of the gauge symmetry \cite{maas2019a}. Rather, it can only be classified according to the global \U[1] symmetry. Thus, there can only be either singlets or non-singlets. Furthermore, because of the group structure -- and similarly to QCD --  any non-singlet gauge-invariant state carries at least three times the \U[1] charge of the scalar field. Thus, besides the absence of degeneracies, the absence of fractional \U[1] charges in the spectrum are unique consequences of requiring gauge symmetry\footnote{There is a long history about the (im)possibility of constructing gauge-invariant fractional charges in QCD, which essentially applies verbatim to the present case: see \cite{Lavelle:1995ty} for an overview.}.

In the uncharged sector the FMS mechanism can be used in a straightforward way to determine the spectrum \cite{maas2019}. This is shown in figure \ref{fig:fun_spectra}, which shows a scalar of the same mass as the elementary Higgs and a vector particle with the same mass as the singlet elementary gauge boson. The latter is not a coincidence, but can be traced back to the breaking pattern in the BEH mechanism \cite{sondenheimer2020}. In particular, the theory is gapped, in contrast to the perturbative result. In the usual GUT scenarios, gapping is relegated to presumed strong interactions in the unbroken subsector \cite{Langacker:1980js}. Exploratory lattice investigations in the past \cite{maas2018} supported the FMS mechanism result. These also showed no hint for residual strong interactions within systematic uncertainties.

The situation is more involved in the charged sector\footnote{Note that we do not consider the implications of a broken \U[1] symmetry, which is even more involved.}. Bose statistics together with the group theory of \SU[3] imply that the simplest operators in the charged sector, at least for scalars and vectors, do not reduce to single-particle operators at leading order of the FMS expansion, but involve more fields \cite{maas2019a}. This is also true for other uncharged channels, and in that case the emerging states are potentially scattering states \cite{maas2019a,sondenheimer2020}. However, since the \U[1] charge is conserved, the lightest state in this channel is necessarily stable, and cannot be a scattering state. A possibility is to interpret the result in a constituent particle model, similar to the quark model. This would imply that the charged scalar and vector should be degenerate in mass, and have twice the mass of non-singlet gauge bosons, i.\ e.\ $2m_a$ \cite{maas2019a}. This is also shown in figure \ref{fig:fun_spectra}. First lattice results suggest that, at least in the vector channel, this may indeed be adequate \cite{maas2018}.

\subsection{The spectrum on the lattice}

The spectrum of the theory is thus currently the primary interesting objective. The exploratory investigations \cite{maas2017,maas2018} only covered a very limited set of quantum number channels. We expand this to all singly charged and uncharged channels with any $J^{P}$ up to $J=2$. In addition, the uncharged channels have defined charge parity $C$, and we consider all possibilities here as well. Considering only $J\le 2$ is based on the prejudice that the mass increases with increasing $J$, as it does in QED and QCD. Moreover, experiments usually only search for particles with $J\le 2$, i.\ e.\ up to graviton-like particles, and hence these are also the experimentally most interesting channels. Given that we expect most relevant features to be applicable to realistic theories, this appears a suitable setting.

Constructing operators in this theory is quite similar as in QCD. We can have either pure gauge boson operators, akin to glueballs, or operators involving the scalar field. The latter are hadron-like and can be divided into meson-like operators involving scalars and antiscalars or baryon-like operators with multiples of three scalars (or antiscalars). Only the latter can carry a non-vanishing \U[1] charge. Since glueball-like operators have shown no tendency to have appreciable overlap with light states, we concentrate on hadron-like operators.

Generalizing from the QCD case in \cite{dudek2010}, we consider the hadron-like gauge-invariant operators
\begin{align}
    O^M_{\mu_1,...\mu_N}&=\scF^\dagger D_{\mu_N}...D_{\mu_1}\scF\\
    O^B_{\{\mu_i\},\{\nu_i\},\{\rho_i\}}&=\epsilon_{abc}\qty(D_{\mu_{N_\mu}}...D_{\mu_1}\scF)_a\qty(D_{\nu_{N_\nu}}...D_{\nu_1}\scF)_b\qty(D_{\rho_{N_\rho}}...D_{\rho_1}\scF)_c
\end{align}
with $N\le 3$ and $N_\mu+N_\nu+N_\rho\le 3$. The operators $O^M$ are uncharged and the operators $O^B$ are charged. Operators of defined $J^{P(C)}$ are then obtained from a Clebsch-Gordan construction. Subduction is then used to create defined lattice representations of the operators. In addition, we included a couple of further operators in some channels, which are generalized from \cite{wurtz2015,maas2018,maas2015}. In some channels these operators do not provide a non-vanishing result. Here, we constructed scattering operators from these operators, assuming that they will eventually provide also sufficient overlap with the ground-state. Details will be presented elsewhere \cite{dobson:up}.

\begin{figure}[h]
    \includegraphics[width=\textwidth,trim={0 -0.5cm 0 0},clip]{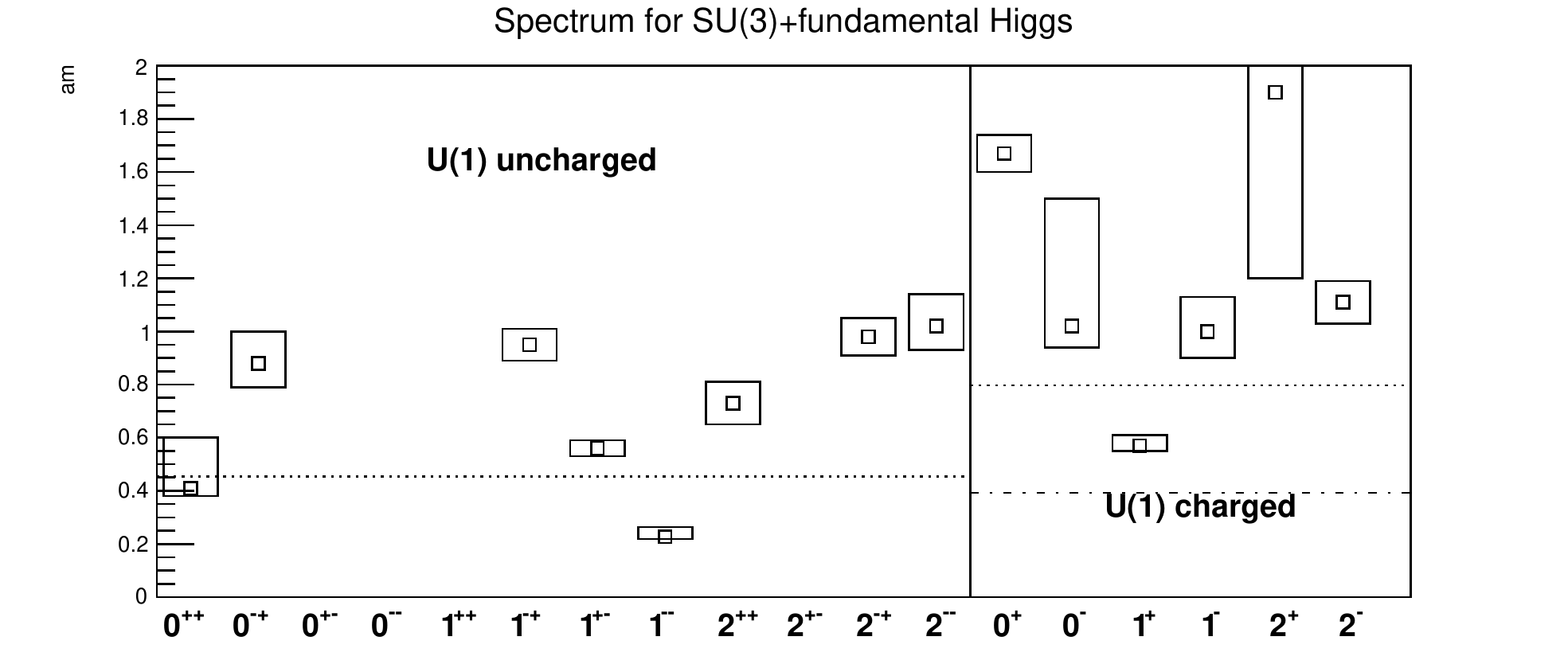}
	\caption{The spectrum on the lattice at a fixed volume of $16^4$ at $\beta=8.433600$, $\kappa=0.488003$, and $\gamma=9.544000$. If no box is present no statistically significant signal was yet observed in the channel. The dotted line on the left-hand side gives the first scattering state from the lightest particle in the uncharged sector, and on the right-hand side for the lightest charged state together with an uncharged vector. The dashed-dotted line gives the prediction for the charged vectors and scalars.}
	\label{fig:fun_spectra_lattice}
\end{figure}

The implementation of the operators is straightforward. We enlarge the operator basis by using smearing as in \cite{maas2018}, through a combination of stout smearing for the links and APE smearing for the scalar fields. In many channels the correlators are quite noisy. Still, in many channels suitable plateaus can be identified and masses extracted. A preliminary result of the spectrum at a fixed volume of $16^4$ is shown in figure \ref{fig:fun_spectra_lattice}. Where available, the results agree with \cite{maas2018}, which is an independent verification as both codes have been developed separately.

It is visible that the lightest uncharged state is the vector, following the same systematics as previously observed \cite{maas2018,maas2015}. The uncharged scalar is compatible with a pure scattering state, and for every other channel with a statistically significant signal the states are even heavier.

In general we observed the $\U[1]$-charged states to be heavier than uncharged ones, which is consistent with expectations from FMS for theories of this kind. The lightest state identified is an axial vector, which had previously not been considered in the spectroscopy and the FMS approach. While this state is still in qualitative agreement with the broad picture of the perturbative FMS expansion predictions, it is also substantially heavier than the leading order constituent model. The vector, which has been previously observed to be in agreement with the prediction \cite{maas2018}, is even heavier. Since the charged states show very strong volume dependence \cite{maas2018}, the systematic control needs to be improved before any conclusions can be drawn.

\section{The adjoint case}

In the adjoint case for $N\ge 3$ a new feature arises, the existence of multiple possible breaking patterns. This has a peculiar consequence. At tree-level in perturbation theory, it is possible to realize different patterns by choice of the vacuum expectation value. While this may change at higher orders \cite{kajantie1998}, this indicates a certain arbitrariness. It is not clear, whether this holds true beyond perturbation theory \cite{maas2019a}. However, this is decisive, as the FMS prediction for the spectrum qualitatively depends on which pattern is realisable. That appears odd at first sight, as this seems to be gauge-dependent. As will be seen below, this is not the case, in a quite non-trivial way. Still, this implies that understanding this non-perturbatively is required.

To this end, we specialize here to $N=3$. In this case our potential has the general form 
\begin{equation}\label{eqn:adj_potential}
    V = -\mu^2\tr\qty[\scA^2] + \lambda\tr\qty[\scA^2]^2
\end{equation}
if the global \ZZ[2] symmetry is not explicitly broken. For $N>3$, the potential gets more involved \cite{maas2019a}.

It is important to note here that symmetry breaking on the lattice implies that the relevant configurations show long-range order per configuration. To identify the breaking thus requires to identify the ordering on a configuration-by-configuration basis by suitable observables.

\subsection{Global symmetry}\label{sec:adj_phases}

Physically, the only relevant symmetry\footnote{There is formally also the \ZZ[3] symmetry of the gauge sector. As it acts trivially on the continuum theory, it does not play a dynamical role in this limit, and is therefore ignored.} is the global \ZZ[2] symmetry, as it encodes gauge-invariant physics. A suitable gauge-invariant order parameter is given by 
\begin{equation}\label{eqn:adj_z2_order}
    \mathcal{O}_{\ZZ[2]}=\ev{\qty[\frac{1}{V}\sum_x \frac{\det\scA(x)}{|\det\scA(x)|}]^2}.
\end{equation}
On a finite lattice, this never vanishes. Thus, to determine its fate it is necessary to extrapolate it to infinite volume in actual calculations. Then, the symmetry is unbroken only when the quantity becomes zero in the infinite-volume limit. Otherwise it is spontaneously broken.

Note, however, that without explicit symmetry breaking the path integral on the lattice still sums for every field configuration $\qty{\scA}$ also over the one $\qty{-\scA}$, and thus $\ev{\scA}=0$ is necessarily true. Therefore an order parameter like \cref{eqn:adj_z2_order} is needed.

\begin{figure}[t!]
    \centering
    \begin{tikzpicture}
        \node (fig) at (0,0) {\includegraphics[width=0.9\textwidth,trim={2.5cm 1cm 1cm 1cm},clip]{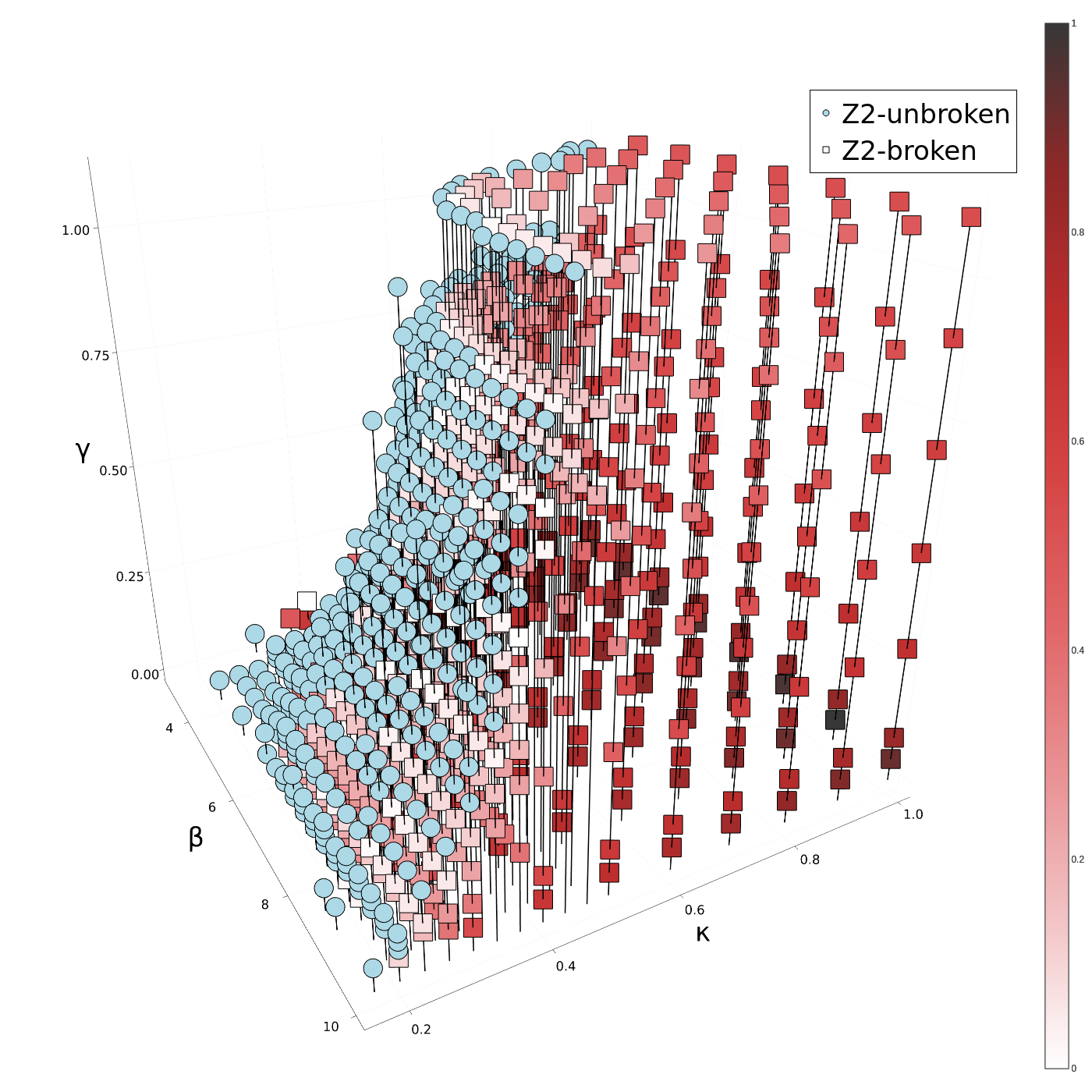}};
        
        \node[font=\large\sffamily, anchor=base] (SU2) at (fig.47) {$\SU[2]\times\U[1]$};
        \node[font=\large\sffamily, anchor=north, yshift=0.2cm] (U1) at (fig.-47) {$\U[1]\times\U[1]^{*}$};
        \node[font=\Large\sffamily, anchor=north, rotate=90,yshift=1.2cm] (cbartitle) at (fig.east) {Breaking angle/pattern};
        \node[font=\Large\sffamily, anchor=north, rotate=90] (cbartitle) at (fig.east) {$\U[1]\times\U[1]$};
    \end{tikzpicture}
    \caption{The non-perturbative phase diagram of the $\SU[3]+\text{adj.}$ theory shows a clear separation into a \ZZ[2]-broken and -unbroken region. These regions coincide also with regions where the gauge-symmetry is un-/broken, see section \ref{sec:local}. In the broken region we observe only one general breaking pattern. In the limiting cases close to the phase transition and the low-$\gamma$/high-$\kappa$ region, the breaking angle approaches slowly the special breaking patterns.}
    \label{fig:adj_phasediag}
\end{figure}
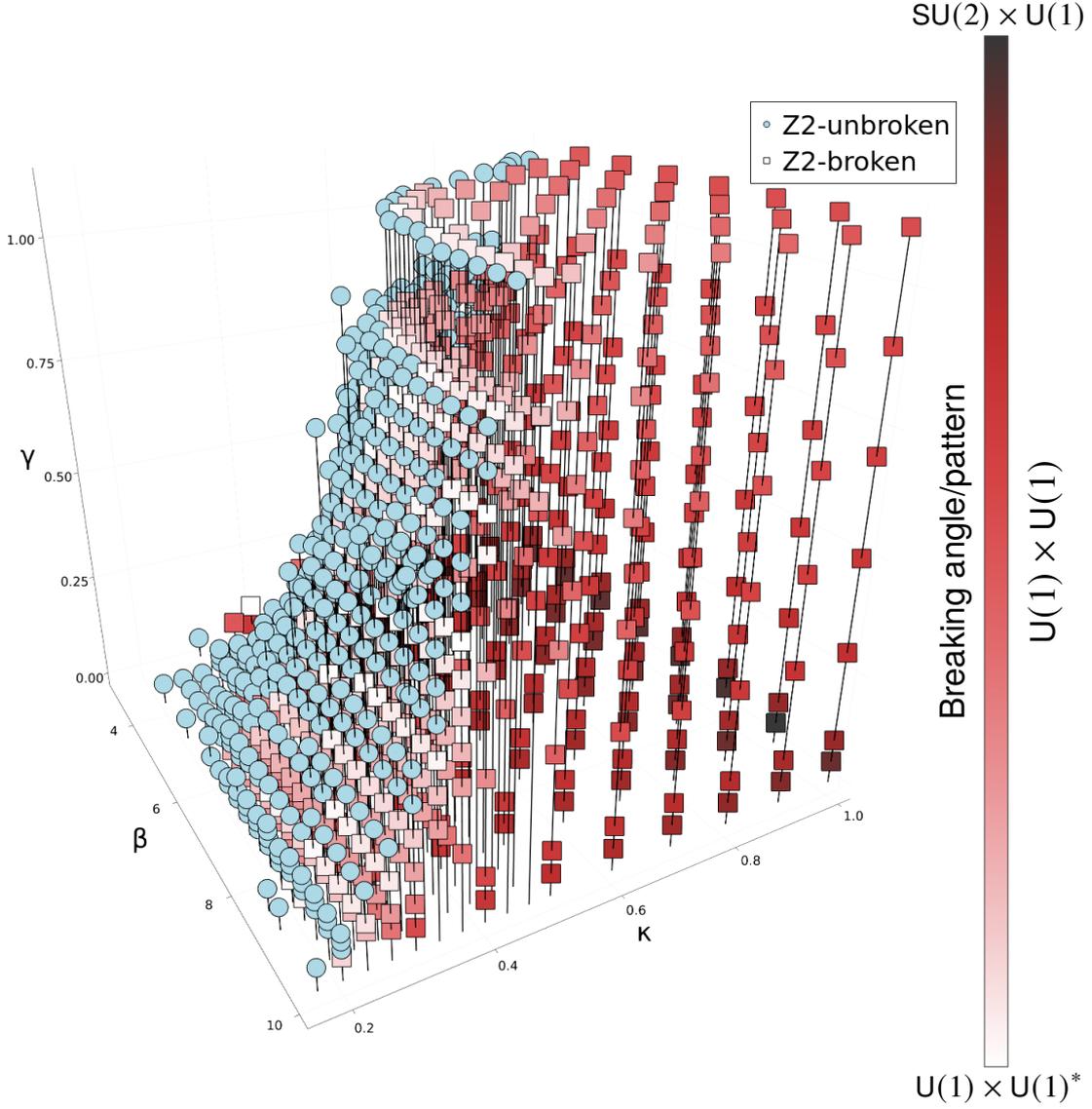

The result for the phase diagram is shown in figure \ref{fig:adj_phasediag}. It is clearly visible that the phase diagram separates into two phases, one with and one without spontaneous breaking of the \ZZ[2] symmetry.

\subsection{Local symmetry}\label{sec:local}

We start by following the path of the standard BEH mechanism. Therefore the scalar field is split again into a non-vanishing vacuum expectation value (vev) $w>0$ pointing into a certain direction $\scA_0$ and fluctuations around it
\begin{equation}\label{eqn:adj_beh}
    \scA\qty(x) = \ev{\scA} + \sigma\qty(x) = w\scA_0 + \sigma\qty(x)
\end{equation}
Note that by specifying $w\scA_0$ we completely fix the direction of the vev. This has far-reaching consequences to be discussed below. For perturbation theory as well as the FMS mechanism to work it is necessary that fluctuations around $w\scA_0$ are sufficiently small.

The difference compared to the fundamental case is that more than one breaking pattern is possible \cite{oraifeartaigh1986}, i.\ e.\ there is more than one unitarily inequivalent choice for $w\scA_0$. Since $\scA_0$ is diagonalisable and traceless, all possible directions of the vev are parametrized by an angle $\theta_0$ in the two-dimensional Cartan of $\SU[3]$, with
\begin{align}\label{eqn:adj_breaking_angle}
    \scA_0 = \cos(\theta_0) T_3 + \sin(\theta_0)T_8\,.
\end{align}
Depending on $\theta_0$, $\scA_0$ can have either two or three distinct non-zero eigenvalues, giving a breaking pattern of $\SU[2]\times\U[1]$ or $\U[1]\times\U[1]$ \cite{maas2019a}. \Cref{tab:adj_patterns} lists the possible patterns. The choice of breaking pattern is thus equivalent to the choice of the angle $\theta_0$. 

\begin{table}[t!]
	\centering
	\begin{tabular}{c|c|c|c}
		& $\SU[2]\times\U[1]$ & $\U[1]\times\U[1]^{*}$ & $\U[1]\times\U[1]$ \\\hline\hline
        eigenvalues\textsu{$^\ast$} 
        	& $\{\lambda,\lambda, -2\lambda\}$ & $\{\lambda, -\lambda, 0\}$ & $\{\lambda_1, \lambda_2, -(\lambda_1+\lambda_2)\}$ \\\hline
		\multirow{2}{*}[-3pt]{vev-alignment $\qty(\scA_0^3,\scA_0^8)$} & $\pm\qty(0,1)$ & $\pm\qty(1,0)$ & \multirow{2}{*}[-3pt]{other} \\
		& $\pm\qty(\sqrt{3}/2,\pm1/2)$ &  $\pm\qty(1/2,\pm\sqrt{3}/2)$ & \\\hline
		\rule{0pt}{15pt}breaking angle $\theta_0$ & $\frac{\qty(2n+1)\pi}{6} \text{ for }n=0,\dots,5$& $\frac{2n\pi}{6} \text{ for }n=0,\dots,5$ & other
		\\\bottomrule
		\multicolumn{4}{l}{\footnotesize{
            \textsu{$^\ast$} up to permutation, and where $\lambda_i$ are nonzero and distinct (i.e. $\lambda_1\neq\lambda_2$).
        }}
	\end{tabular}
	\caption{Symmetry breaking patterns depend on the direction of the chosen vev, $\theta_0$. In the general case there are two distinct nonzero parameters defining the eigenvalues and the vev is invariant under $\U[1]\times\U[1]$ transformations. For twelve special values of $\theta_0$ the parameters are (anti-)degenerate, leading either to an $\SU[2]\times\U[1]$ symmetry or to a special case of the $\U[1]\times\U[1]$ pattern, denoted by $\U[1]\times\U[1]^{*}$.
	}
    \label{tab:adj_patterns}
\end{table}

To implement the condition in \cref{eqn:adj_beh} requires gauge-fixing, which leads to some subtleties. Consider for the moment unitary gauge, which locally diagonalizes $\scA(x)$. Here it is possible to impose a strong ordering, in the sense that the diagonal elements are sorted by size. As a consequence the vacuum expectation value
\begin{align}\label{eqn:vev}
    \ev{\frac{1}{V}\sum_x \scA(x)}
\end{align}
is necessarily non-vanishing if $\scA(x)$ is non-zero in any sizeable fraction of the volume. Long-range order is thereby enforced on the scalar field, and it appears as if a BEH effect is present throughout the phase diagram. This kind of gauge-fixing can be implemented by locally diagonalizing all $\scA(x)$, which is numerically straightforward. We have done this, and confirmed this statement in our simulations.

The reason is, of course, a trade-off when gauge-fixing. By enforcing order on the scalar field, any disorder is transferred to the gauge fields, which therefore will strongly fluctuate. Thus, while a vacuum condensate arises in this way, the FMS mechanism may not work, as the gauge field fluctuations invalidate perturbation theory.

There are (at least) two options to avoid imposing long-range order on the scalar field: the results tested in unitary and Landau--'t Hooft gauges both coincide.
The first possibility is as follows: instead of enforcing a strong ordering of the eigenvalues, we note that the breaking patterns as listed in \cref{tab:adj_patterns} do not rely on the ordering of the eigenvalues.
However, long-range ordering implies that the ordering of the eigenvalues need to be correlated over long distances. 
Thus, we fix to unitary gauge, but only admit gauge transformation from the coset $\SU[3]/S_3$, where $S_3$ is the permutation group of order three. In practice, diagonalization algorithms do not ensure such an ordering. To enforce it, we determine before diagonalization an angle $\theta$ by taking the trace of the scalar field normalized to unit determinant with $T_8$, and then extract an angle in analogy to \cref{eqn:adj_breaking_angle}. We then reorder the eigenvalues after diagonalization such that for the gauge-fixed scalar field, which has the normalized form \cref{eqn:adj_breaking_angle}, the resulting angle $\theta_0$ coincides with the original one within an interval of $\pi/6$ within the sectors listed in table \ref{tab:adj_patterns}.

Of course, keeping the order free implies again that for every configuration there exists another configuration with exactly opposite ordering, due to the \ZZ[2] symmetry. Thus, in this case the vev in \cref{eqn:vev} vanishes. The detection of the long-range order is then possible using \cite{caudy2008,maas2019}
\begin{align}\label{eqn:adj_gauge_order_lat}
    \ev{ \qty(\frac{1}{V}\sum_x \scA(x))^2}.
\end{align}
This allows to unambiguously identify the presence of long-range ordering. But perturbation theory, and the FMS mechanism, rely on a unique value for the vacuum expectation value. Thus, to eventually obtain configurations, which are fixed to the same gauge as in the continuum, requires, after establishing that for a given set of lattice parameter long-range ordering persists, the eigenvalues to agree with \cref{eqn:adj_beh}. Thus, this is a two-step process on the lattice.

The other possibility is to use Landau--'t Hooft gauge \cite{maas2017}. Since Landau gauge enforces maximum smoothness on the gauge fields locally \cite{cucchieri1996} this largely prevents local correlations being transferred to the gauge fields. As a consequence, it is not possible to obtain a non-vanishing vacuum expectation value everywhere, but only where long-range ordering exists. Consequently, we find in the numerical simulations that the results in Landau-'t Hooft gauge agree with the ones from gauge-fixing to unitary gauge restricted to $\SU[3]/S_3$. Again, only after fully fixing the ordering of the eigenvalues, now for the spatially averaged vev, it makes sense to compare to continuum results.

Finally, we also want to know which breaking pattern is realised for a given set of parameters. In any finite calculation no true degeneracies will arise for the eigenvalues, so formally every result looks like the $\U[1]\times\U[1]$ pattern. Only when approaching an infinite system it can approach either of the other patterns. To check this, we calculate the breaking angle $\theta_0$ as given in \cref{eqn:adj_breaking_angle} and then extrapolate. This quantity can be obtained either directly from the gauge-fixed scalar field or indirectly from gauge invariant quantities like $\sum_x\tr\qty[\scA\qty(x)^3]$ and $\sum_x\tr\qty[\scA\qty(x)^2]$. It turned out that both ways yield the same results within errors. Furthermore it needs to be mentioned that the breaking patterns are periodic in terms of the breaking angles and therefore it is sufficient to restrict the angle to the range $\theta_0\in\qty[0,\pi/6]$, where the lower end is the $\U[1]\times\U[1]^{*}$-pattern and the upper end gives the $\SU[2]\times\U[1]$ pattern.

\begin{figure}[t!]
    \centering
    \begin{subfigure}[b]{0.45\textwidth}
        \includegraphics[width=\textwidth,trim={0 -0.5cm 0 0},clip]{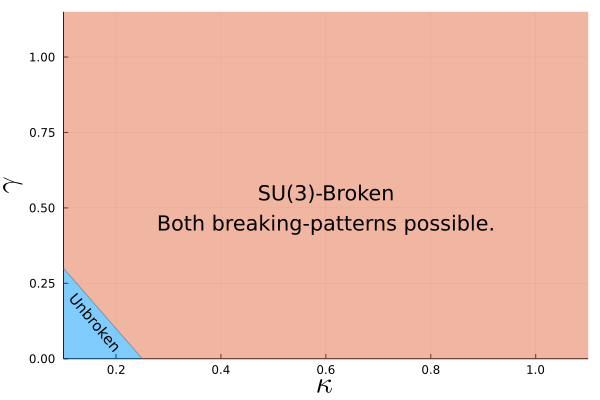}
        \caption{perturbative tree-level phase diagram}
        \label{fig:adj_pt_phasediag}
    \end{subfigure}
    \begin{subfigure}[b]{0.45\textwidth}
        \centering
        \begin{tikzpicture}
            \node (fig) at (0,0) {\includegraphics[width=0.93\textwidth,trim={0 0 1.5cm 0},clip]{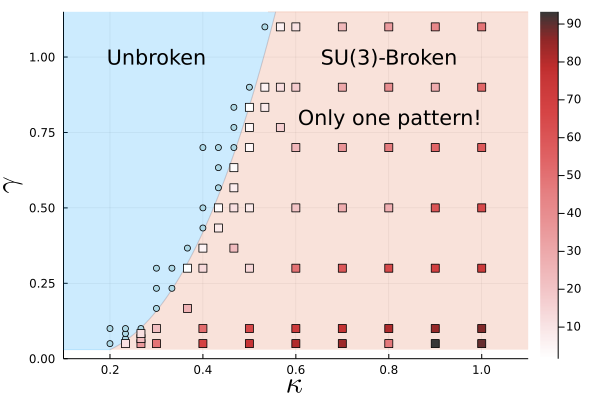}};
            \node[font=\tiny\sffamily, anchor=base] (SU2) at (fig.35) {$\SU[2]\times\U[1]$};
            \node[font=\tiny\sffamily, anchor=north] (U1) at (fig.-28) {$\U[1]\times\U[1]^{*}$};
            \node[font=\scriptsize\sffamily, anchor=north, rotate=90, yshift=0.1cm] (cbartitle) at (fig.east) {Breaking angle/pattern};
            \node[font=\scriptsize\sffamily, anchor=north, rotate=90, yshift=-0.3cm] (cbartitle) at (fig.east) {$\U[1]\times\U[1]$};
        \end{tikzpicture}
        \caption{non-perturbative phase diagram ($\beta=6$)}
        \label{fig:adj_gf_phasediag}
    \end{subfigure}
    \caption{The $\beta$-independent perturbative tree-level phase diagram (left) and the non-perturbative phase diagram (right) differ in the location of phases and the observed breaking patterns.}
    \label{fig:adj_comp_phasediag}
\end{figure}
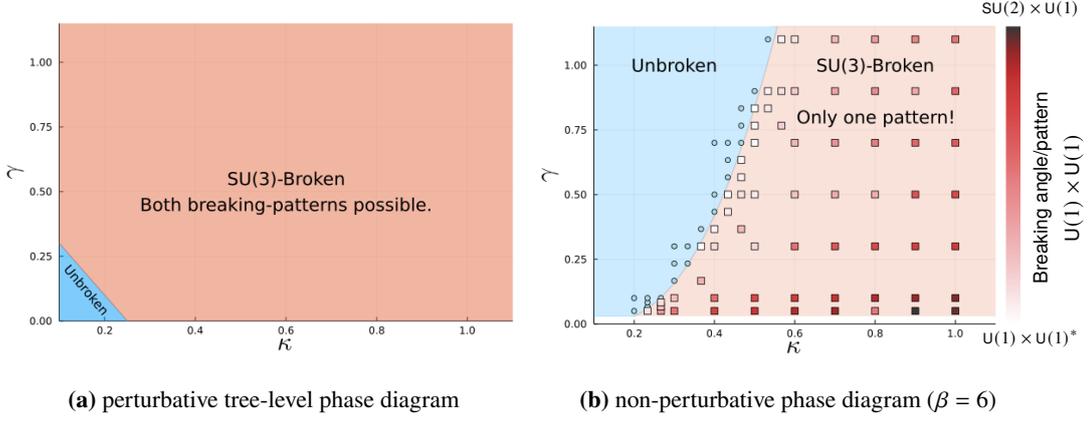

In \cref{fig:adj_comp_phasediag} the result for the phase diagram is shown for a specific value of $\beta$ compared with the tree-level prediction from perturbation theory. It can immediately be seen that already the general separation into the \SU[3]-broken and \SU[3]-unbroken regions differ quite substantially. For simplicity only a few points for the non-perturbative phase diagram in the unbroken region are shown, but the whole blue shaded area has been checked carefully. The second quite substantial difference is that only one breaking pattern is realised which which happens to be the general $\U[1]\times\U[1]$ case.

Extending this kind of analysis to different values of the gauge coupling $\beta$ allows us to get an impression of the full non-perturbative phase diagram in the parameter space. In \cref{fig:adj_phasediag} the so-obtained phase diagram is shown. The two different phases have been obtained from the gauge invariant \ZZ[2]-breaking order parameter \cref{eqn:adj_z2_order} and from the gauge-symmetry breaking order parameter \cref{eqn:adj_gauge_order_lat} in a fixed gauge. In all of the simulated points in the parameter space we observed a one-to-one correspondence between \ZZ[2]- and gauge-symmetry breaking. Apart from that we also observe a clear separation into the ordered and disordered phases.

Again, for the full phase diagram we only observed the general $\U[1]\times\U[1]$ breaking pattern, but one can make two observations regarding the breaking angle based on \cref{fig:adj_phasediag,fig:adj_gf_phasediag}. Near the phase transition the breaking angle is always very small (indicated by almost white rectangles) and it continuously increases when going deeper into the BEH-like region. The other observation to be made is by going to larger $\kappa$- and smaller $\gamma$-values the breaking angles approach $\pi/6$ which corresponds to the $\SU[2]\times\U[1]$ breaking pattern. What we want to emphasize here is that we have not observed either of the special $\U[1]\times\U[1]^{*}$- or the $\SU[2]\times\U[1]$-breaking pattern. However, the gradient of the breaking angle indicates that in the limiting cases, i.e. at the phase transition and in the low-$\gamma$/high-$\kappa$ region, the special patterns may be realised.

The apparent connection to the \ZZ[2] phases can be now understood in the following way. Because its order parameter \cref{eqn:adj_z2_order} depends only on the sign of the determinant, any information about long-range order in the scalar field is lost, as this depends on the relative ordering of the eigenvalues. However, the $\U[1]\times\U[1]^{*}$ case implies a vanishing determinant, due to one zero eigenvalue. Thus, in case of a long-range ordering of this type, \ZZ[2] symmetry would be necessarily intact.

Conversely, for the $\SU[2]\times\U[1]$ case the determinant is also long-range ordered. Thus, this pattern can only be realised if \ZZ[2] is broken. Finally, in the $\U[1]\times\U[1]$ case the determinant depends non-linearly on the relative sign of the two eigenvalues. If one of the eigenvalues is relatively large and the other one is asymmetrically distributed around zero, the determinant can fluctuate wildly in configurations, but there is still long-range order possible for the eigenvalues. Hence, in total there is no way to decide a-priori from the status of the \ZZ[2] symmetry whether a BEH effect is present, at most a subset of the the possible patterns can be deduced, if one is present. Thus, it is highly non-trivial that a correspondence is found.

Finally, it should be remarked that for NLO estimates in three dimensions only the $\SU[2]\times\U[1]$ pattern \cite{kajantie1998} has been found. However, this occurred at very small values of $\gamma$, where we also see that this pattern may emerge\footnote{Towards extremely small values of $\gamma$ acceptance rates drop dramatically, so our algorithm cannot penetrate exactly as deep into this region as was possible in three dimensions.}.

From these results it can be inferred that we have a distinct and unique possible breaking pattern to implement a BEH effect. We can thus uniquely identify the type of predictions of the FMS mechanism applying, and are thus at a useful starting point to test them.

\section{Conclusions and outlook}

We have extended previous investigations of prototypes of GUTs using a manifestly gauge-invariant setup on the lattice. These are important steps to establish an augmentation of perturbation theory to perform successful and manifestly gauge-invariant phenomenology for the wide range of parameter sets not accessible by lattice simulations alone.

On the one hand, we have started to substantially extend the number of quantum number channels investigated for the fundamental case. The patterns we observe are in agreement with those expected in the uncharged sector. However, the charged sector offers some surprises. More systematic investigations will be necessary to establish a firm picture and to fully understand how to use the FMS mechanism to predict reliably the charged spectrum. This will be a next step \cite{dobson:up}.

For the adjoint case we could clarify the situation with multiple breaking patterns. We find that a unique pattern prevails non-perturbatively. While the identification of the pattern is so far tied to lattice methods, this allows to uniquely provide FMS-mechanism predictions, disentangling previous ambiguities \cite{maas2019a}. Further systematics will ultimately tell under which conditions breaking patterns of any type can be realized \cite{dobson:up}.

\acknowledgments

The computational results presented have been obtained using the Vienna Scientific Cluster (VSC) and the HPC center at the University of Graz. The plots in \cref{fig:adj_comp_phasediag,fig:adj_phasediag} have been created using the Plots package \cite{juliaplots2021} of the Julia Programming Language \cite{julia2017}. The ROOT-Framework \cite{rootv62402} has been used in this work. E. D. and B. R. have been supported by the Austrian Science Fund FWF, grant P32760. We are grateful to Vincenzo Afferrante, René Sondenheimer, and Pascal T\"orek for useful discussions.

\bibliographystyle{JHEP}
\bibliography{refs}

\end{document}